\documentstyle[prd,aps,preprint,tighten,epsfig]{revtex}

\begin{document}

\draft

\title{Can the Lepton Flavor Mixing Matrix Be Symmetric?}
\author{\bf Zhi-zhong Xing}
\address{CCAST (World Laboratory), P.O. Box 8730, Beijing 100080, China \\
and \\
Institute of High Energy Physics, P.O. Box 918 (4), 
Beijing 100039, China 
\footnote{Mailing address} \\
({\it Electronic address: xingzz@mail.ihep.ac.cn}) }
\maketitle

\begin{abstract}
Current neutrino oscillation data indicate that the $3\times 3$ lepton 
flavor mixing matrix $V$ is likely to be symmetric about its 
$V_{e3}$-$V_{\mu 2}$-$V_{\tau 1}$ axis. This off-diagonal symmetry 
corresponds to three pairs of {\it congruent} unitarity triangles 
in the complex plane. Terrestrial matter effects can substantially
modify the genuine $CP$-violating parameter and off-diagonal
asymmetries of $V$ in realistic long-baseline experiments of neutrino
oscillations.
\end{abstract}

\pacs{PACS number(s): 14.60.Pq, 13.10.+q, 25.30.Pt} 

\newpage

\section{Introduction}

The observed anomalies of atmospheric \cite{ATM} and solar \cite{SUN}
neutrinos strongly suggest that neutrinos be massive and lepton
flavors be mixed. In the framework of three charged leptons 
and three active neutrinos, 
the phenomena of flavor mixing and $CP$ violation are described by a
unitary matrix $V$, which relates the neutrino mass eigenstates
$(\nu_1, \nu_2, \nu_3)$ to the neutrino flavor eigenstates
$(\nu_e, \nu_\mu, \nu_\tau)$:
\begin{equation}
\left ( \matrix{
\nu_e \cr
\nu_\mu \cr
\nu_\tau \cr} \right )
= \left ( \matrix{
V_{e1}	& V_{e2} & V_{e3} \cr
V_{\mu 1} & V_{\mu 2} & V_{\mu 3} \cr
V_{\tau 1} & V_{\tau 2} & V_{\tau 3} \cr} \right ) 
\left ( \matrix{
\nu_1 \cr
\nu_2 \cr
\nu_3 \cr} \right ) .
\end{equation}
The unitarity of $V$ represents two sets of normalization and
orthogonality conditions:
\begin{eqnarray}
\sum_i \left ( V_{\alpha i} V^*_{\beta i} \right )
& = & ~ \delta_{\alpha\beta} \; , 
\nonumber \\
\sum_\alpha \left ( V_{\alpha i} V^*_{\alpha j} \right )
& = & ~ \delta_{ij} \; ,
\end{eqnarray}
where Greek and Latin subscripts run over $(e, \mu, \tau)$ and
$(1,2,3)$, respectively.
If neutrinos are Dirac particles, a full parametrization of $V$
requires four independent parameters -- 
three mixing angles and one $CP$-violating phase, for example.
If neutrinos are Majorana particles, however, two additional 
$CP$-violating phases need be introduced for a complete parametrization 
of $V$. In both cases, $CP$ and $T$ violation in normal neutrino
oscillations depends only upon a single rephasing-invariant parameter
${\cal J}$ \cite{Jarlskog}, defined through
\begin{equation}
{\rm Im} \left ( V_{\alpha i} V_{\beta j} V^*_{\alpha j}
V^*_{\beta i} \right ) = {\cal J} \sum_{\gamma, k}
\left (\epsilon_{\alpha\beta\gamma} \epsilon_{ijk} \right ) \; ,
\end{equation}
where $(\alpha, \beta, \gamma)$ and $(i,j,k)$ run respectively
over $(e, \mu, \tau)$ and $(1,2,3)$. A major goal of the future long-baseline
neutrino oscillation experiments is to measure $|V_{\alpha i}|$ and 
${\cal J}$ as precisely as possible \cite{LBL}. Once the matrix elements
of $V$ are determined to a good degree of accuracy, a stringent test of 
its unitarity will become available.

As a straightforward consequence of the unitarity of $V$, two 
interesting relations can be derived from the normalization conditions
in Eq. (2):
\begin{eqnarray}
|V_{e 2}|^2 - |V_{\mu 1}|^2  & = & |V_{\mu 3}|^2 - |V_{\tau 2}|^2
\nonumber \\
& = & |V_{\tau 1}|^2 - |V_{e 3}|^2 \equiv \Delta_{\rm L} \; ,
\end{eqnarray}
and
\begin{eqnarray}
|V_{e 2}|^2 - |V_{\mu 3}|^2 & = & |V_{\mu 1}|^2 - |V_{\tau 2}|^2
\nonumber \\
& = & |V_{\tau 3}|^2 - |V_{e 1}|^2 \equiv \Delta_{\rm R} \; .
\end{eqnarray}
The off-diagonal asymmetries $\Delta_{\rm L}$ and $\Delta_{\rm R}$ characterize
the geometrical structure of $V$ about its $V_{e1}$-$V_{\mu 2}$-$V_{\tau 3}$
and $V_{e 3}$-$V_{\mu 2}$-$V_{\tau 1}$ axes, respectively. If
$\Delta_{\rm L} = 0$ held, $V$ would be symmetric about the
$V_{e1}$-$V_{\mu 2}$-$V_{\tau 3}$ axis. Indeed the counterpart of 
$\Delta_{\rm L}$ in the quark sector is very small 
(of order $10^{-5}$ \cite{Xing95}); i.e., the $3\times 3$ quark mixing 
matrix is almost symmetric about its $V_{ud}$-$V_{cs}$-$V_{tb}$ axis.
An exactly symmetric flavor mixing matrix may hint at an underlying
flavor symmetry, from which some deeper understanding of the fermion mass 
texture can be achieved \cite{FXreview}. In this sense, the tiny 
off-diagonal asymmetry of the quark flavor mixing matrix is likely to arise
from a slight breakdown of certain flavor symmetries of quark mass matrices.

The purpose of this paper is to examine whether the lepton flavor mixing 
matrix $V$ is really symmetric or not. In section II, we find that current 
neutrino oscillation data strongly favor $\Delta_{\rm R} =0$; i.e., 
$V$ is possible to be symmetric about its $V_{e 3}$-$V_{\mu 2}$-$V_{\tau 1}$ 
axis. It remains too early to get any phenomenological constraints on 
$\Delta_{\rm L}$, unless very special assumptions are made. In section III,
we point out that the off-diagonal symmetry $\Delta_{\rm R}=0$ corresponds 
to three pairs of {\it congruent} unitarity triangles in the complex plane. 
Taken realistic long-baseline experiments of neutrino oscillations into
account, the terrestrial matter effects on ${\cal J}$, $\Delta_{\rm L}$ and 
$\Delta_{\rm R}$ are briefly discussed in section IV. 
Section V is devoted to some further discussions about possible
implications of $\Delta_{\rm R}=0$ on specific textures of lepton mass 
matrices. Finally we summarize our main results in section VI.

\section{Off-diagonal Symmetry}

Current experimental data \cite{ATM,SUN} strongly favor the hypothesis that
atmospheric and solar neutrino oscillations are dominated by 
$\nu_\mu \rightarrow \nu_\tau$ and $\nu_e \rightarrow \nu_\mu$ transitions, 
respectively. Thus their mixing factors
$\sin^2 2\theta_{\rm atm}$ and $\sin^2 2\theta_{\rm sun}$ have rather simple
relations with the elements of the lepton flavor mixing matrix $V$.
The mixing factor associated with the CHOOZ (or Palo Verde) reactor 
neutrino oscillation experiment \cite{CHOOZ}, denoted as 
$\sin^2 2\theta_{\rm chz}$, is also a simple function of $|V_{\alpha i}|$ 
in the same hypothesis. 
The explicit expressions of $\sin^2 2\theta_{\rm sun}$,
$\sin^2 2\theta_{\rm atm}$ and $\sin^2 2\theta_{\rm chz}$ read as
follows:
\begin{eqnarray}
\sin^2 2\theta_{\rm sun} & = & 4 |V_{e1}|^2 |V_{e2}|^2 \; ,
\nonumber \\
\sin^2 2\theta_{\rm atm} & = & 4 |V_{\mu 3}|^2 
\left ( 1 - |V_{\mu 3}|^2 \right ) \; ,
\nonumber \\ 
\sin^2 2\theta_{\rm chz} & = & 4 |V_{e3}|^2 
\left ( 1 - |V_{e3}|^2 \right ) \; .
\end{eqnarray}
An analysis of the Super-Kamiokande data on atmospheric neutrino
oscillations \cite {ATM} yields $0.88 \leq \sin^2 2\theta_{\rm atm} \leq 1.0$ 
and $1.6 \times 10^{-3} ~ {\rm eV}^2 \leq \Delta m^2_{\rm atm} 
\leq 4.0 \times 10^{-3} ~ {\rm eV}^2$ at the $90\%$ confidence level. 
Corresponding to $\Delta m^2_{\rm chz} > 2.0 \times 10^{-3} ~ {\rm eV}^2$,
$\sin^2 2\theta_{\rm chz} < 0.18$ can be drawn from the CHOOZ 
experiment \cite{CHOOZ}. We restrict ourselves to the large-angle
Mikheyev-Smirnov-Wolfenstein (MSW) solution to the solar neutrino
problem \cite{MSW}, as it gives the best global fit of present data. At the
$99\%$ confidence level, $0.56 \leq \sin^2 2\theta_{\rm sun} \leq 0.99$ 
and $2.0 \times 10^{-5} ~ {\rm eV}^2 \leq \Delta m^2_{\rm sun} \leq
5.0 \times 10^{-4} ~ {\rm eV}^2$ have been obtained \cite{LMA}.
 
With the help of Eqs. (2) and (6), one may express 
$|V_{e1}|^2$, $|V_{e2}|^2$, $|V_{e3}|^2$ and $|V_{\mu 3}|^2$ in terms
of $\theta_{\rm sun}$, $\theta_{\rm atm}$ and $\theta_{\rm chz}$:
\begin{eqnarray}
|V_{e1}|^2 & = & \frac{1}{2} \left ( \cos^2\theta_{\rm chz}
\pm \sqrt{\cos^4\theta_{\rm chz} - \sin^2 2\theta_{\rm sun}} \right ) \; ,
\nonumber \\
|V_{e2}|^2 & = & \frac{1}{2} \left ( \cos^2\theta_{\rm chz}
\mp \sqrt{\cos^4\theta_{\rm chz} - \sin^2 2\theta_{\rm sun}} \right ) \; ,
\nonumber \\
|V_{e3}|^2 & = & \sin^2 \theta_{\rm chz} 
~ {\rm or} ~ \cos^2 \theta_{\rm chz} \; ,
\nonumber \\
|V_{\mu 3}|^2 & = & \sin^2 \theta_{\rm atm} 
~ {\rm or} ~ \cos^2 \theta_{\rm atm} \; .
\end{eqnarray}
Without loss of generality, three mixing angles ($\theta_{\rm sun}$, 
$\theta_{\rm atm}$ and $\theta_{\rm chz}$) can all be arranged to lie
in the first quadrant. Then we need only adopt the solution
$|V_{e3}|^2 = \sin^2 \theta_{\rm chz}$ \cite{Xing02}, in accord with 
$\sin^2 2\theta_{\rm chz} < 0.18$. We may also express
$|V_{\tau 3}|^2$ in terms of $\theta_{\rm atm}$
and $\theta_{\rm chz}$, once the normalization relation 
$|V_{e3}|^2 + |V_{\mu 3}|^2 + |V_{\tau 3}|^2 =1$ is taken into account.
It turns out that useful experimental constraints are achievable for 
those matrix elements in the first row and in the third column of $V$.
However, it is impossible to get any constraints on the other four 
matrix elements of $V$, unless some special assumptions are made
\footnote{Allowing the Dirac-type $CP$-violating phase of $V$ to vary between
0 and $\pi$, Fukugita and Tanimoto \cite{Tanimoto} have presented the 
numerical ranges of all nine $|V_{\alpha i}|$ by use of current neutrino 
oscillation data. This rough construction of the lepton flavor mixing matrix 
is actually unable to shed light on its off-diagonal asymmetries and 
$CP$-violating features.}.
This observation means that it remains too early to get any instructive 
information on the off-diagonal asymmetry $\Delta_{\rm L}$ from current 
neutrino oscillation experiments, but it is already possible to examine 
whether $\Delta_{\rm R} =0$ coincides with current data and what its 
implications can be on leptonic $CP$ violation and unitarity triangles.

To see whether $\Delta_{\rm R} =0$ is compatible with the present data of
solar, atmospheric and reactor neutrino oscillations, we simply set
$|V_{e2}|^2 = |V_{\mu 3}|^2$ in Eq. (7) and then obtain
\begin{equation}
\sin^2 2\theta^{(\pm)}_{\rm sun} \; =\; \sin^2 2\theta_{\rm atm} -
\left ( 1 \pm \sqrt{1 - \sin^2 2\theta_{\rm atm}} \right )
\left ( 1 - \sqrt{1 - \sin^2 2\theta_{\rm chz}} \right ) \; .
\end{equation}
As $\sin^2 2\theta_{\rm chz} < 0.18$ \cite{CHOOZ}, the second term on
the right-hand side of Eq. (8) serves as a small correction to the
leading term $\sin^2 2\theta_{\rm atm}$. The difference between
$\sin^2 2\theta^{(+)}_{\rm sun}$ and $\sin^2 2\theta^{(-)}_{\rm sun}$
is therefore insignificant. Indeed $\sin^2 2\theta_{\rm chz} =0$ 
leads definitely to $\sin^2 2\theta^{(+)}_{\rm sun} = 
\sin^2 2\theta^{(-)}_{\rm sun} = \sin^2 2\theta_{\rm atm}$, 
as a straightfoward result of $\Delta_{\rm R} =0$. Allowing 
$\sin^2 2\theta_{\rm atm}$ to vary in the experimental range
$0.88 \leq \sin^2 2\theta_{\rm atm} \leq 1.0$,
we plot the numerical dependence of $\sin^2 2\theta^{(\pm)}_{\rm sun}$ on 
$\sin^2 2\theta_{\rm chz}$ in Fig. 1. One can observe that the 
values of $\sin^2 2\theta^{(\pm)}_{\rm sun}$ predicted from Eq. (8) are 
consistent very well with current experimental data. Thus
we conclude that a vanishing or tiny off-diagonal asymmetry of $V$ about
its $V_{e3}$-$V_{\mu 2}$-$V_{\tau 1}$ axis is strongly favored.

\section{Unitarity triangles}

Let us proceed to discuss possible implications of $\Delta_{\rm R} =0$ on 
the leptonic unitarity triangles. It is known that six orthogonality 
relations of $V$ in Eq. (2) correspond to six triangles in the complex 
plane \cite{FXreview}, as illustrated
in Fig. 2. These six triangles totally have eighteen different sides and nine 
different inner angles. Unitarity requires that all six triangles have the
same area amounting to ${\cal J}/2$, where ${\cal J}$ is just the 
rephasing-invariant measure of leptonic $CP$ violation defined in Eq. (3). 

Now that the off-diagonal asymmetries $\Delta_{\rm L}$ and $\Delta_{\rm R}$
describe the geometrical structure of $V$, they must have direct relations with
the unitarity triangles in the complex plane. Indeed it is easy to show that
$\Delta_{\rm L} =0$ or $\Delta_{\rm R} =0$ corresponds to the congruence 
between two unitarity triangles; i.e.,
\begin{eqnarray}
\Delta_{\rm L} = 0  & ~~~~ \Longrightarrow ~~~~ & 
\triangle_e \; \cong \; \triangle_1 \; , 
\nonumber \\
&& \triangle_\mu \; \cong \; \triangle_2 \; ,
\nonumber \\
&& \triangle_\tau \; \cong \; \triangle_3 \; ; 
\end{eqnarray}
and
\begin{eqnarray}
\Delta_{\rm R} = 0  & ~~~~ \Longrightarrow ~~~~ & 
\triangle_e \; \cong \; \triangle_3 \; , 
\nonumber \\
&& \triangle_\mu \; \cong \; \triangle_2 \; ,
\nonumber \\
&& \triangle_\tau \; \cong \; \triangle_1 \; .
\end{eqnarray}
As $\Delta_{\rm R} =0$ is expected to be rather close to reality, we
draw the conclusion that the unitarity triangles $\triangle_e$
and $\triangle_3$ must be approximately congruent with each other.
A similar conclusion can be drawn for the unitarity triangles
$\triangle_\mu$ and $\triangle_2$ as well as $\triangle_\tau$ and 
$\triangle_1$. The long-baseline experiments of neutrino oscillations 
in the near future will tell whether an approximate congruence exists
between $\triangle_e$ and $\triangle_1$ or between $\triangle_\tau$ and
$\triangle_3$. A particularly interesting possibility would be
$\Delta_{\rm L} \approx \Delta_{\rm R} \approx 0$; i.e., only two of
the six unitarity triangles are essentially distinct.

Next we examine how large the area of each unitarity triangle 
(i.e., ${\cal J}/2$) can maximally be in the limit $\Delta_{\rm R} =0$, 
in which $V$ is parametrized as follows:
\begin{equation}
V \; = \; \left ( \matrix{ 
c_x c_z & s_x c_z & s_z \cr
- c_x s_x s_z - c_x s_x e^{-i\delta} &
- s^2_x s_z + c^2_x e^{-i\delta} &
s_x c_z \cr 
- c^2_x s_z + s^2_x e^{-i\delta} & 
- c_x s_x s_z - c_x s_x e^{-i\delta} & 
c_x c_z \cr } \right ) 
\left ( \matrix{
e^{i\rho}	& 0	& 0 \cr
0	& e^{i\sigma}	& 0 \cr
0	& 0	& 1 \cr} \right ) \; ,
\end{equation}
where $s_x \equiv \sin\theta_x$, $c_z \equiv \cos\theta_z$, and so on.
The merit of this phase choice is that the Dirac-type $CP$-violating
phase $\delta$ does not appear in the effective mass term of the neutrinoless 
double beta decay \cite{FX01}, which depends only upon the Majorana phases 
$\rho$ and $\sigma$. Without loss of generality, one may arrange the 
mixing angles $\theta_x$ and $\theta_z$ to lie in the first
quadrant. Three $CP$-violating phases ($\delta, \rho, \sigma$) can take
arbitrary values from 0 to $2\pi$. Clearly
${\cal J} = c^2_x s^2_x c^2_z s_z \sin\delta$ holds. With the help of
Eq. (6) or (7), we are able to figure out the relations between 
$(\theta_x, \theta_z)$ and $(\theta_{\rm sun}, \theta_{\rm chz})$. The 
result is 
\begin{eqnarray}
\sin^2 2\theta_x & = & \frac{4 \sin^2 2\theta_{\rm sun}}
{\left ( 1 + \sqrt{1 - \sin^2 2\theta_{\rm chz}} \right )^2} \; ,
\nonumber \\
\sin^2 2\theta_z & = & \sin^2 2\theta_{\rm chz} \; .
\end{eqnarray}
Then we obtain
\begin{equation}
{\cal J}_{(\pm)} \; =\; \frac{\sqrt{2}}{4} ~
\sin^2 2\theta^{(\pm)}_{\rm sun} \sin\delta ~
\frac{\sqrt{1- \sqrt{1 - \sin^2 2\theta_{\rm chz}}}}
{1 + \sqrt{1 - \sin^2 2\theta_{\rm chz}}} \; ,
\end{equation}
where $\sin^2 2\theta^{(\pm)}_{\rm sun}$ has been given in Eq. (8). 
Again the difference between ${\cal J}_{(+)}$ and ${\cal J}_{(-)}$
is insignificant. If $\delta =0$ or $\pi$ held, we would arrive at 
${\cal J}_{(\pm)} =0$. In general, however, $CP$ symmetry is expected to 
break down in the lepton sector. For illustration, we plot the numerical 
dependence of ${\cal J}_{(\pm)}/\sin\delta$ 
on $\sin^2 2\theta_{\rm chz}$ in Fig. 3, where the experimentally allowed
values of $\sin^2 2\theta_{\rm atm}$ are used. It is obvious that the
upper bound of ${\cal J}_{(\pm)}$ (when $\delta = \pi/2$) can be as large as 
a few percent, only if $\sin^2 2\theta_{\rm chz} \geq 0.01$. This result 
implies that leptonic $CP$ and $T$ violation might be observable in the future
long-baseline neutrino oscillation experiments.

\section{Matter effects}

In realistic long-baseline experiments of neutrino oscillations, the
terrestrial matter effects must be taken into account \cite{Barger}. The 
pattern of neutrino oscillations in matter can be expressed in the same form
as that in vacuum, however, if we define the {\it effective} neutrino
masses $\tilde{m}_i$ and the {\it effective} lepton flavor mixing matrix
$\tilde{V}$ in which the terrestrial matter effects are already 
included \cite{Xing00}. Note that $\tilde{m}_i$ are functions of
$m_j$, $|V_{ej}|$ and $A$; and $\tilde{V}_{\alpha i}$ are functions of
$m_j$, $V_{\beta j}$ and $A$, where $A$ is the matter parameter and its
magnitude depends upon the neutrino beam energy $E$ and the background density
of electrons $N_e$. The analytically exact relations between
$(\tilde{m}_i, \tilde{V}_{\alpha i})$ and $(m_i, V_{\alpha i})$ can be
found in Ref. \cite{Xing00}, if $N_e$ is assumed to be a constant. 
In analogy to the definitions of 
${\cal J}$, $\Delta_{\rm L}$ and $\Delta_{\rm R}$, 
the {\it effective} $CP$-violating parameter 
$\tilde{\cal J}$ and off-diagonal asymmetries $\tilde{\Delta}_{\rm L}$ and 
$\tilde{\Delta}_{\rm R}$ can be defined as follows:
\begin{equation}
{\rm Im} \left ( \tilde{V}_{\alpha i} \tilde{V}_{\beta j} 
\tilde{V}^*_{\alpha j} \tilde{V}^*_{\beta i} \right ) = 
\tilde{\cal J} \sum_{\gamma, k}
\left (\epsilon_{\alpha\beta\gamma} \epsilon_{ijk} \right ) \; ,
\end{equation}
where $(\alpha, \beta, \gamma)$ and $(i,j,k)$ run respectively
over $(e, \mu, \tau)$ and $(1,2,3)$; 
\begin{eqnarray}
|\tilde{V}_{e 2}|^2 - |\tilde{V}_{\mu 1}|^2  & = & 
|\tilde{V}_{\mu 3}|^2 - |\tilde{V}_{\tau 2}|^2
\nonumber \\
& = & |\tilde{V}_{\tau 1}|^2 - |\tilde{V}_{e 3}|^2 \equiv 
\tilde{\Delta}_{\rm L} \; ,
\end{eqnarray}
and
\begin{eqnarray}
|\tilde{V}_{e 2}|^2 - |\tilde{V}_{\mu 3}|^2 & = & 
|\tilde{V}_{\mu 1}|^2 - |\tilde{V}_{\tau 2}|^2
\nonumber \\
& = & |\tilde{V}_{\tau 3}|^2 - |\tilde{V}_{e 1}|^2 \equiv 
\tilde{\Delta}_{\rm R} \; .
\end{eqnarray}
It is then interesting to examine possible departures of $(\tilde{\cal J},
\tilde{\Delta}_{\rm L}, \tilde{\Delta}_{\rm R})$ from
$({\cal J}, \Delta_{\rm L}, \Delta_{\rm R})$ in a concrete 
experimental scenario.

To illustrate, we compute $\tilde{\cal J}$, $\tilde{\Delta}_{\rm L}$ and
$\tilde{\Delta}_{\rm R}$ using the typical inputs $\theta_x = 40^\circ$,
$\theta_z = 5^\circ$ and $\delta = 90^\circ$, which yield
${\cal J} = 0.021$, $\Delta_{\rm L} = 0.166$ and $\Delta_{\rm R} =0$ in
vacuum. The neutrino mass-squared differences are taken to be 
$\Delta m^2_{21} = \Delta m^2_{\rm sun} = 5\times 10^{-5} ~ {\rm eV}^2$
and $\Delta m^2_{31} = \Delta m^2_{\rm atm} = 3\times 10^{-3} ~ {\rm eV}^2$.
The explicit formulas relevant to our calculation have been given in
Ref. \cite{Xing00}. 
We plot the numerical dependence of $\tilde{\cal J}/{\cal J}$,
$\tilde{\Delta}_{\rm L}$ and $\tilde{\Delta}_{\rm R}$ on the matter
parameter $A$ in Fig. 4, where both the cases of neutrinos 
($+A$) and antineutrinos ($-A$) are taken into account. One can see
that the off-diagonal symmetry $\Delta_{\rm R}=0$ in vacuum can
substantially be spoiled by terrestrial matter effects. The deviation
of $\tilde{\Delta}_{\rm L}$ from $\Delta_{\rm L}$ and that of
$\tilde{\cal J}$ from ${\cal J}$ are remarkably large,
if $A > 10^{-5} ~ {\rm eV}^2$. 

As emphasized in Ref. \cite{Xing01}, there exists the simple reversibility
between the fundamental neutrino mixing parameters in vacuum and their
effective counterparts in matter. The former can therefore be expressed
in terms of the latter, allowing more straightforward extraction of the
genuine lepton mixing quantities (including ${\cal J}$, $\Delta_{\rm L}$ and
$\Delta_{\rm R}$) from a variety of long-baseline neutrino oscillation
experiments. If $|V_{\alpha i}|$ are determined to a very high degree of
accuracy, it will be possible to test the unitarity of $V$ \cite{FXreview}
and establish leptonic $CP$ violation through the non-zero area of six 
unitarity triangles \cite{Smirnov} even in the absence of a direct 
measurement of ${\cal J}$.

\section{Further discussions}

If $\Delta_{\rm R}=0$ really holds, 
one may wonder whether this off-diagonal symmetry of $V$
hints at very special textures of the neutrino
mass matrix $M_\nu$ and (or) the charged lepton mass matrix $M_l$. 
In the following, we take two simple but
instructive examples to illustrate possible implications of 
$\Delta_{\rm R}=0$ on $M_l$ and $M_\nu$.

\begin{center}
{\bf Example A}
\end{center}

In no conflict with current data on atmospheric \cite{ATM},
solar \cite{SUN} and reactor \cite{CHOOZ} neutrino oscillations, 
a remarkably simplified form of $V$ with $\Delta_{\rm R} =0$
and $\delta =\rho =\sigma =0$ is
\begin{equation}
{\bf V} \; =\; \left ( \matrix{
\displaystyle\frac{c}{\sqrt{2}}	& 
\displaystyle\frac{c}{\sqrt{2}}	& 
s \cr\cr
- \displaystyle\frac{1+s}{2}	&
\displaystyle\frac{1-s}{2}	& 
\displaystyle\frac{c}{\sqrt{2}} \cr\cr
\displaystyle\frac{1-s}{2}	& 
- \displaystyle\frac{1+s}{2}	& 
\displaystyle\frac{c}{\sqrt{2}} \cr} \right ) \; ,
\end{equation}
where $s\equiv \sin\theta \ll 1$ and 
$c\equiv \cos\theta \approx 1$ \cite{X00}. 
In the flavor basis where $M_l$ is diagonal, $M_\nu$ can be given as
\begin{eqnarray}
M_\nu & = & {\bf V} \left ( \matrix{
m_1	& 0	& 0 \cr
0	& m_2	& 0 \cr
0	& 0	& m_3 \cr} \right ) {\bf V}^{\rm T}
\nonumber \\
& = & m_1 {\bf N}_1 ~ + ~ m_2 {\bf N}_2 ~ + ~ m_3 {\bf N}_3 \; ,
\end{eqnarray}
where symmetric matrices ${\bf N}_1$, ${\bf N}_2$ and ${\bf N}_3$ read
\begin{eqnarray}
{\bf N}_1 & = & \left ( \matrix{
\displaystyle\frac{c^2}{2} 	&
- \displaystyle\frac{c (1+s)}{2\sqrt{2}}	&
\displaystyle\frac{c (1-s)}{2\sqrt{2}} \cr\cr
	&
\displaystyle\frac{(1+s)^2}{4}	&
- \displaystyle\frac{1-s^2}{4} \cr\cr
	&
	&
\displaystyle\frac{(1-s)^2}{4} \cr} \right ) \; ,
\nonumber \\
\nonumber \\
{\bf N}_2 & = & \left ( \matrix{
\displaystyle\frac{c^2}{2} 	&
\displaystyle\frac{c (1-s)}{2\sqrt{2}}	&
- \displaystyle\frac{c (1+s)}{2\sqrt{2}} \cr\cr
	&
\displaystyle\frac{(1-s)^2}{4}	&
- \displaystyle\frac{1-s^2}{4} \cr\cr
	&
	&
\displaystyle\frac{(1+s)^2}{4} \cr} \right ) \; ,
\nonumber \\
\nonumber \\
{\bf N}_3 & = & \left ( \matrix{
s^2 	&
~~~~~ \displaystyle\frac{cs}{\sqrt{2}} ~~~~~	&
~~~ \displaystyle\frac{cs}{\sqrt{2}} ~~ \cr\cr
	&
\displaystyle\frac{c^2}{2}	&
\displaystyle\frac{c^2}{2} \cr\cr
	&
	&
\displaystyle\frac{c^2}{2} \cr} \right ) \; .
\end{eqnarray}
The texture of $M_\nu$ is rather complicated, hence it is
difficult to observe any hidden flavor symmetry associated 
with lepton mass matrices. 

\begin{center}
{\bf Example B}
\end{center}

The result of $M_\nu$ in Example A will become simpler, if
$s =0$ is further taken (i.e., $\bf V$ is of the bi-maximal mixing 
form \cite{Barger2}).
In this special case, however, simple textures of $M_l$ and $M_\nu$
can be written out in a more general flavor basis.
It is easy to show that $M_l$ and $M_\nu$ of the
following textures lead to $\bf V$ with $s=0$:
\begin{eqnarray}
M_l & = & \frac{C_l~}{2} \left [ \left ( \matrix{
0	& 0	& 0 \cr
0	& 1	& 1 \cr
0	& 1	& 1 \cr} \right ) + \left ( \matrix{
\delta_l & 0	& 0 \cr
0	& 0	& \varepsilon_l \cr
0	& \varepsilon_l & 0 \cr} \right ) \right ] \; , 
\nonumber \\
M_\nu & = & \frac{C_\nu}{2} \left [ \left ( \matrix{
1	& 0	& 0 \cr
0	& 1	& 0 \cr
0	& 0	& 1 \cr} \right ) + \left ( \matrix{
0 	& \varepsilon_\nu	& 0 \cr
\varepsilon_\nu	& 0	& 0 \cr
0	& 0 & \delta_\nu \cr} \right ) \right ] \; , 
\end{eqnarray}
where $\delta_{l,\nu}$ and $\varepsilon_{l,\nu}$ are small perturbative
parameters \cite{FX96}. 
In the limit $\delta_{l,\nu} =\varepsilon_{l,\nu} =0$, 
$M_l$ has the $S(2)_{\rm L} \times S(2)_{\rm R}$ symmetry and
$M_\nu$ displays the $S(3)$ symmetry \cite{FTY}. 
The perturbative corrections in $M_l$ allow electron and muon to 
acquire their masses:
\begin{equation}
\{ m_e ~, ~ m_\mu ~, ~ m_\tau \} \; =\; \frac{C_l}{2}
\{ |\delta_l| ~, ~ |\varepsilon_l| ~, ~ 2 + \varepsilon_l \} \; .
\end{equation}
We then arrive at $C_l = m_\mu + m_\tau \approx 1.88$ GeV,
$|\varepsilon_l| = 2m_\mu/(m_\mu + m_\tau) \approx 0.11$ and
$|\delta_l| = 2m_e/(m_\mu + m_\tau) \approx 5.4 \times 10^{-4}$.
The perturbative corrections in $M_\nu$ make three neutrino masses
non-degenerate:
\begin{equation}
\{ m_1, m_2, m_3 \} \; =\; C_\nu \{ 1+\varepsilon_\nu,
1-\varepsilon_\nu, 1+\delta_\nu \} \; .
\end{equation}
As a result, $|\varepsilon_\nu|/|\delta_\nu| \approx 
\Delta m^2_{\rm sun}/(2\Delta m^2_{\rm atm}) \sim 10^{-2}$ for the
large-angle MSW solution to the solar neutrino problem.

Examples A and B illustrate how the off-diagonal symmetry of $V$
($\Delta_{\rm R} =0$) can be reproduced from specific textures of 
lepton mass matrices. 

\section{Summary}

In view of current experimental data on solar and atmospheric neutrino
oscillations, we have discussed the geometrical structure of the 
$3\times 3$ lepton flavor mixing matrix $V$. We find that the present data
strongly favor the off-diagonal symmetry of $V$ about its
$V_{e3}$-$V_{\mu 2}$-$V_{\tau 1}$ axis. This symmetry, if really exists, 
will correspond to three pairs of congruent unitarity
triangles in the complex plane. It remains too early to tell whether $V$
is symmetric or not about its $V_{e1}$-$V_{\mu 2}$-$V_{\tau 3}$ axis.
A brief analysis of terrestrial matter effects on the 
universal $CP$-violating parameter and off-diagonal asymmetries of $V$
has also been made. We expect that future long-baseline experiments of
neutrino oscillations can help establish the texture of the
lepton flavor mixing matrix, from which one could get some insights into
the underlying flavor symmetries responsible for the charged lepton and
neutrino mass matrices.

\acknowledgments{This work was supported in part by the
National Natural Science Foundation of China.}

\newpage

\begin{figure}
\vspace{-1.cm}
\epsfig{file=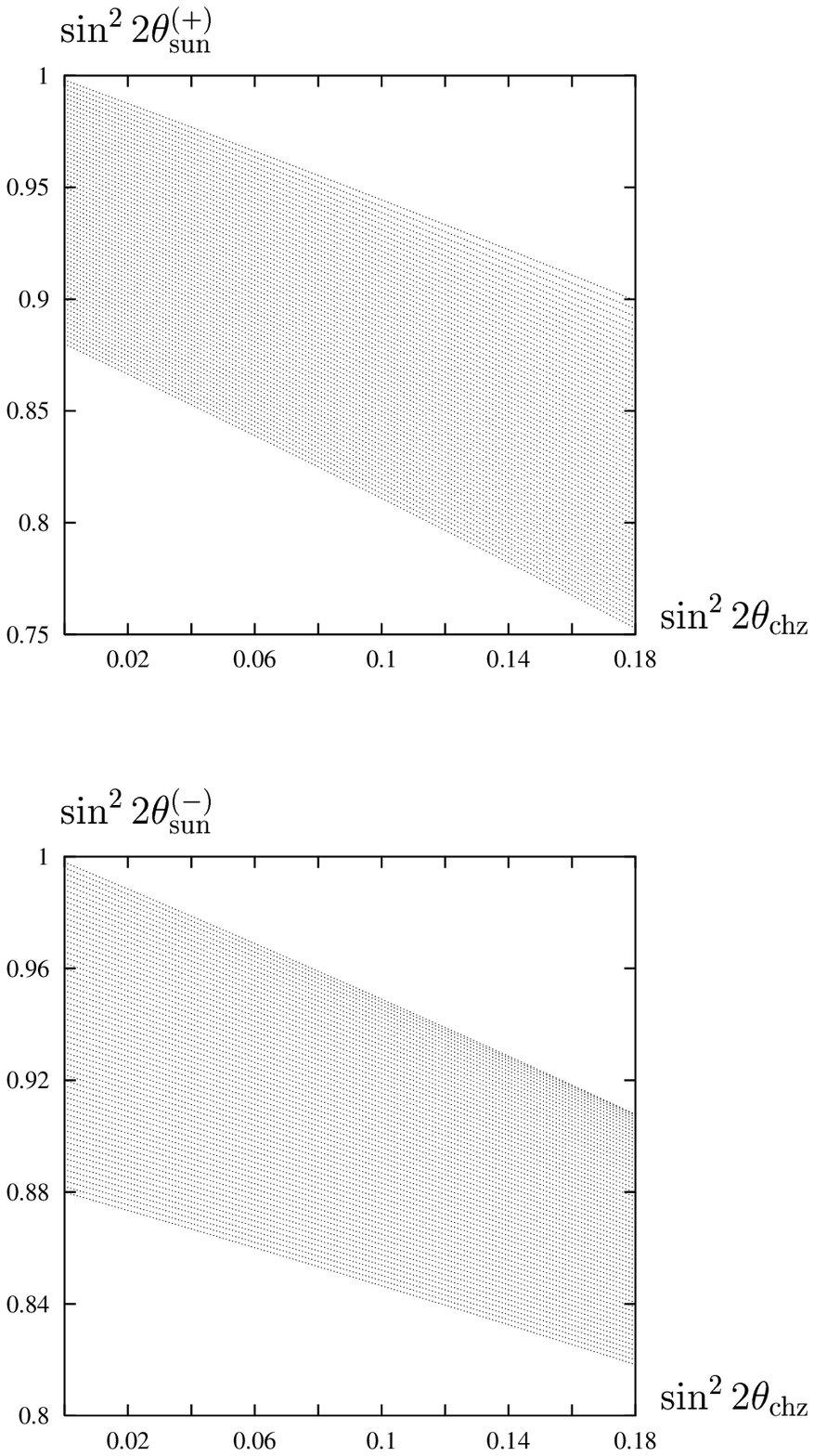,bbllx=1cm,bblly=4cm,bburx=20cm,bbury=32cm,%
width=15cm,height=22cm,angle=0,clip=}
\vspace{-1.5cm}
\caption{Dependence of $\sin^2 2\theta^{(\pm)}_{\rm sun}$ on 
$\sin^2 2\theta_{\rm chz}$, where 
$0.88 \leq \sin^2 2\theta_{\rm atm} \leq 1.0$ has been input.}
\end{figure}

\begin{figure}
\epsfig{file=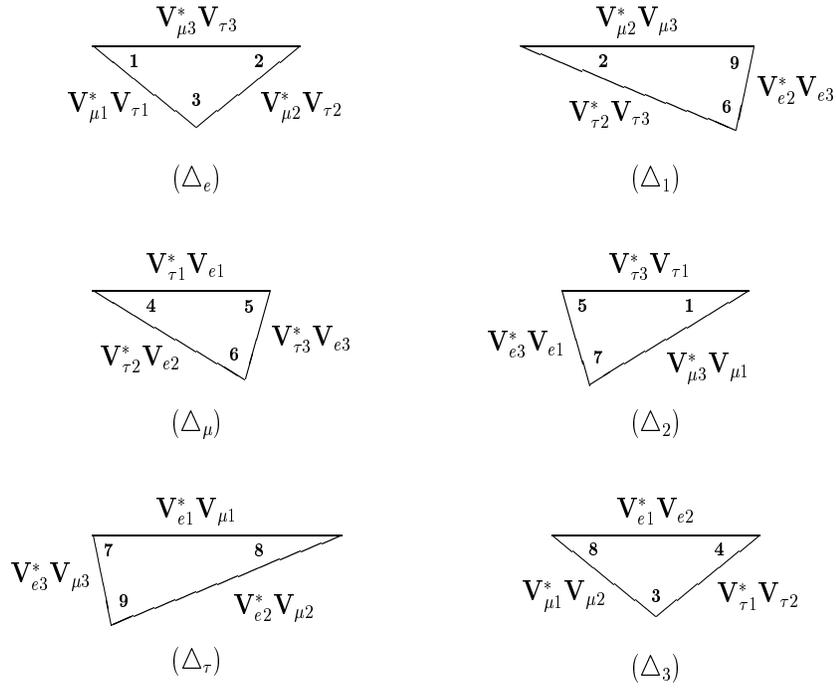,bbllx=1cm,bblly=8cm,bburx=20cm,bbury=32cm,%
width=15cm,height=22cm,angle=0,clip=}
\vspace{-6.7cm}
\caption{Unitarity triangles of the lepton flavor mixing matrix in the
complex plane. Each triangle is named by the Greek or
Latin subscript that does not manifest in its three sides.}
\end{figure}

\begin{figure}
\vspace{-2.2cm}
\epsfig{file=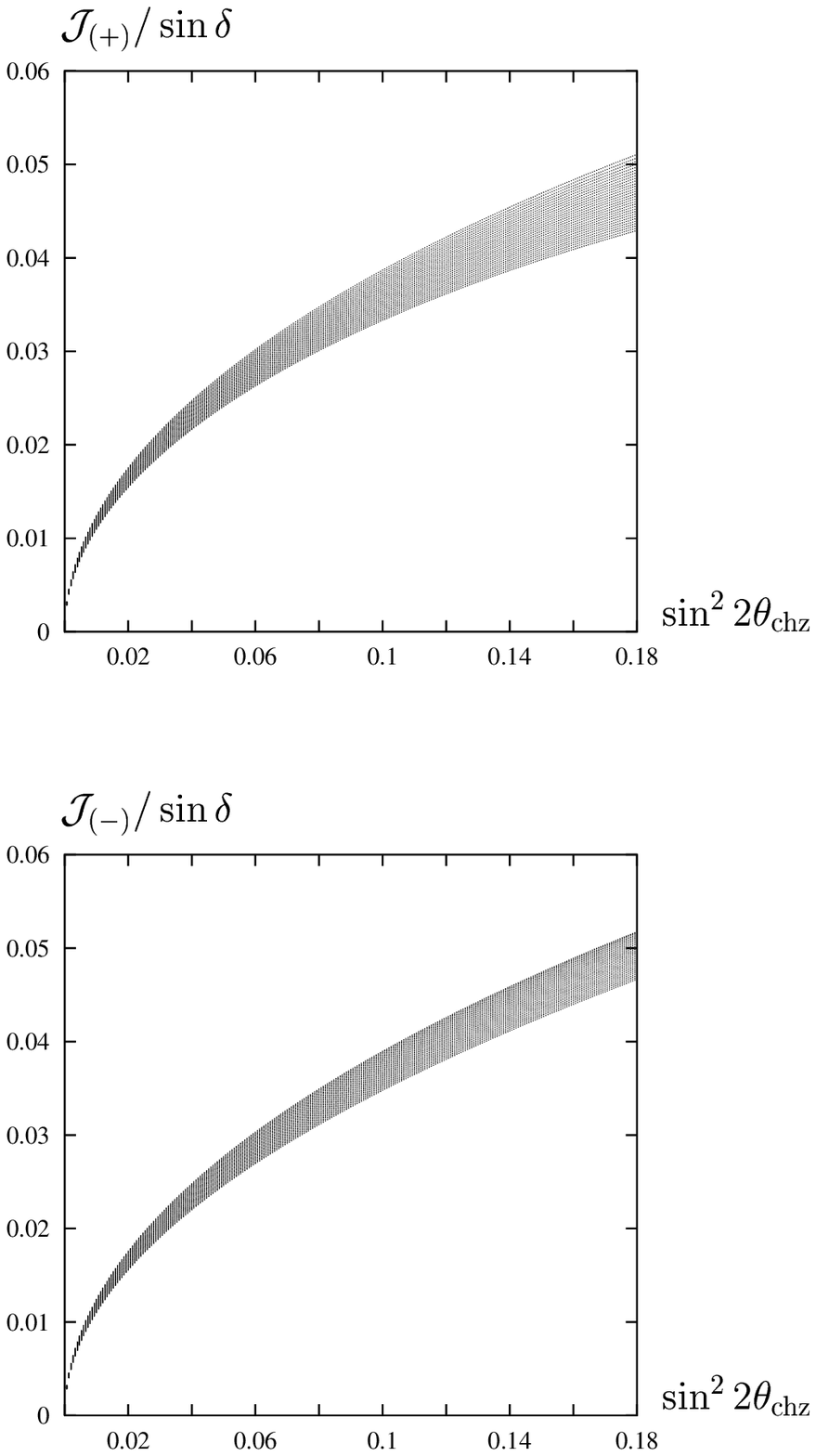,bbllx=1cm,bblly=4cm,bburx=20cm,bbury=32cm,%
width=15cm,height=22cm,angle=0,clip=}
\vspace{-1.5cm}
\caption{Dependence of ${\cal J}_{(\pm)}/\sin\delta$ on 
$\sin^2 2\theta_{\rm chz}$, where 
$0.88 \leq \sin^2 2\theta_{\rm atm} \leq 1.0$ has been input.}
\end{figure}

\begin{figure}
\vspace{-2.2cm}
\epsfig{file=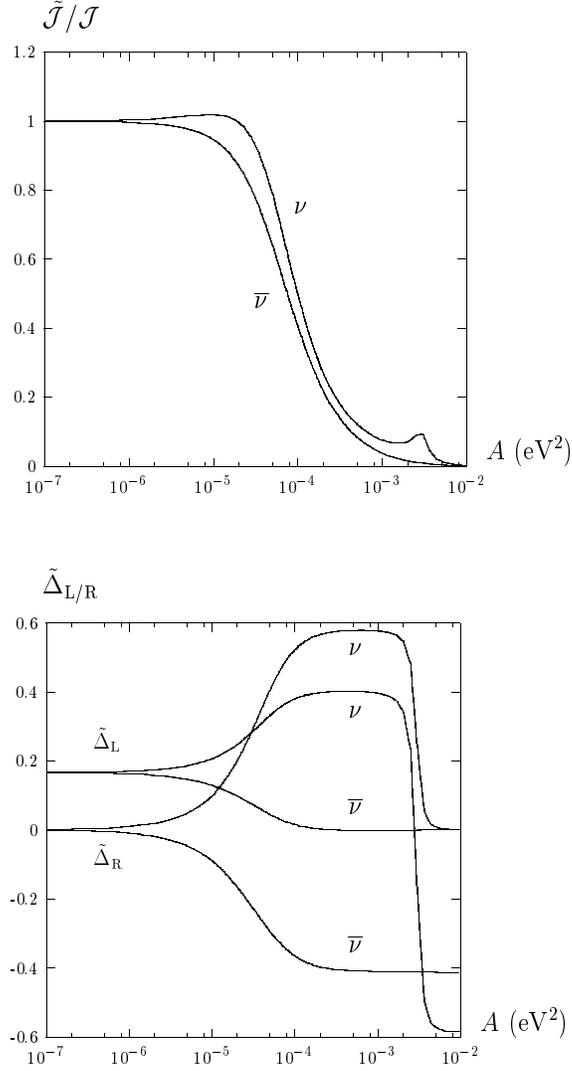,bbllx=1cm,bblly=4cm,bburx=20cm,bbury=32cm,%
width=13cm,height=20cm,angle=0,clip=}
\vspace{-0.2cm}
\caption{Illustrative plots for terrestrial matter effects on
${\cal J}$, $\Delta_{\rm L}$ and $\Delta_{\rm R}$ associated with
neutrinos ($+A$) and antineutrinos ($-A$), where
$\Delta m^2_{21} = 5\times 10^{-5} ~{\rm eV}^2$,
$\Delta m^2_{31} = 3\times 10^{-3} ~{\rm eV}^2$,
$\theta_x = 40^\circ$, $\theta_z = 5^\circ$ and 
$\delta = 90^\circ$ have typically been input.}
\end{figure}

\end{document}